\documentclass[%
 reprint,amsmath,amssymb,aps,prb,floatfix]{revtex4-1}

\usepackage{tgbonum}
\usepackage{graphicx}
\usepackage{dcolumn}
\usepackage{bm}
\usepackage{hyperref} 
\usepackage{times} 
\usepackage{array}
\newcolumntype{P}[1]{>{\centering\arraybackslash}p{#1}}

\usepackage{capt-of}
\usepackage{booktabs}
\usepackage{varwidth}

\usepackage{graphics}
\usepackage{graphicx}
\usepackage{array}
\usepackage{dcolumn}
\usepackage{bm}
\usepackage{float}
\usepackage{enumitem}
\usepackage[colorinlistoftodos, textwidth=4cm, shadow]{todonotes}
\usepackage{amssymb}
\usepackage{hyperref}
\usepackage{multirow}
\usepackage{threeparttable}
\usepackage[export]{adjustbox}
\usepackage{tabularx,booktabs}
\newcolumntype{Y}{>{\centering\arraybackslash}X}
\begin{document}

\title {Tuning flexoelectricty and electronic properties of zig-zag graphene nanoribbons by functionalization}                      

\author{T. Pandey, L. Covaci and F. M. Peeters}
\affiliation{Department of Physics, University of Antwerp, Groenenborgerlaan 171, 2020 Antwerp, Belgium}
\date{\today}

\begin{abstract}
The flexoelectric and electronic properties of zig-zag graphene nanoribbons are explored under mechanical bending using state of the art first principles calculations. A linear dependence of the bending induced out of plane polarization on the applied strain gradient is found. The inferior flexoelectric properties of graphene nanoribbons can be improved by more than two orders of magnitude by hydrogen and fluorine functionalization (CH and CF nanoribbons). A large out of plane flexoelectric effect is predicted for CF nanoribbons. The origin of this enhancement lies in the electro-negativity difference between carbon and fluorine atoms, which breaks the out of plane charge symmetry even for a small strain gradient. The flexoelectric effect can be further improved by co-functionalization with hydrogen and fluorine (CHF Janus-type nanoribbon), where a spontaneous out of plane dipole moment is formed even for flat nanoribbons. We also find that bending can control the charge localization of valence band maxima and therefore enables the tuning of the hole effective masses and band gaps. These results present an important advance towards the understanding of flexoelectric and electronic properties of hydrogen and fluorine functionalized graphene nanoribbons, which can have important implications for flexible electronic applications.
\end{abstract}


\maketitle

\section{Introduction}
The coupling between polarization and strain gradients is an electro-mechanical phenomenon, which can be observed by bending a material. This is known as flexoelectricity, which is present in a variety of materials including soft matter~\cite{deng2014flexoelectricity,porenta2011effect,petrov1989curvature}, liquid crystals~\cite{meyer1969piezoelectric,takahashi1998novel} and crystalline materials~\cite{tagantsev1986piezoelectricity,ma2006flexoelectricity,lu2012mechanical,narvaez2015large,stengel2015flexoelectricity}. The redistribution of charges induces flexoelectricity in materials due to the appearance of strain gradients and it has two microscopic contributions \textemdash~electronic and ionic. The presence of non-uniform strain in a dielectric solid redistributes the ionic charges, leading to the appearance of ionic polarization, whereas the electronic polarization originates from the arrangement of the electronic clouds in the material under applied strain. The dependence of the induced electric polarization upon mechanical deformation of a material is expressed as~\cite{yudin2013fundamentals}:
\begin{equation}
P_{i} =  d_{ijk} \epsilon_{jk} + F_{ijkl}\frac{\partial \epsilon_{jk}}{\partial \epsilon_{k}},
\label{eq1}
\end{equation}
where P$_i$, $\epsilon_{jk}$, $d_{ijk}$ and $f_{ijkl}$ are the components of the polarization, the strain tensor, the piezoelectric tensor and the flexoelectric tensor, respectively.

Following the discovery of flexoelectricity it was found to be a weak effect, which is hardly detectable in bulk materials. Since the induced polarization depends on both the strain gradient and the flexoelectric coefficient; a large gradient is required to generate considerable polarization. A lot of recent experimental efforts have focused on flexoelectricity in perovskite thin films of (Ba/Sr)TiO$_3$~\cite{cross2006flexoelectric, lu2012mechanical} where the strain gradient was introduced during the epitaxial growth ~\cite{li2015giant,ji2010bulk,lee2011giant}. Additionally the recent discovery of strain gradient mediated photovoltaic response in BaFeO$_3$~\cite{chu2015enhancement} has renewed the interest in flexoelectricity. 

Different from bulk materials, layered two-dimensional materials (2D) are able to sustain large strain gradients and are therefore expected to show strong flexoelectricity~\cite{duerloo2013flexural,naumov2009unusual,michel2016static,kundalwal2017strain}. For instance, graphene monolayer is one of the best known elastic materials and it can sustain  more than 10 \% mechanical strains ~\cite{lee2008measurement,zhao2020improving}. Together with these superior elastic properties, the reduced lattice symmetry of 2D materials makes them promising for piezo- and flexoelectric applications~\cite{michel2016static,michel2017piezoelectricity,dong2017large}. For example monolayer MoS$_2$~\cite{dong2017large, wang2019probing} and its Janus structures~\cite{dong2017large,yagmurcukardes2019electronic} exhibit large out of plane piezoelectricity exceeding the most common 3D piezoelectric crystal AlN~\cite{bechmann1958elastic}. Experimentally, the electro-mechanical properties of monolayer MoS$_2$  were explored by perturbing the structure by atomic force microscopy and a large flexoelectric effect induced by the structural deformation was proposed~\cite{abdollahi2019converse}. 

\begin{table*}
\centering
\caption{The monolayer in-plane lattice parameters (\textit{a}, \textit{b}), bond lengths (\textit{d}) and zig-zag nanoribbons width (along the $y$ direction), bending stiffness (C$_b$), buckling height ($\Delta_{bH}$), finite thickness of nanoribbons ($t$) total number of zig-zag chains in nanoribbons (N$_{\mathrm{chain}}$) and out of plane flexoelectric coefficient (F$_{zyzy}$) for C, CH, CF and CHF. The listed lattice parameters correspond to a rectangular unit-cell. The out of plane flexoelectric coefficients are obtained by linear fitting of the induced polarization from Figure~\ref{fig3}.}
\label{Table:1}
\begin{tabular}{c| c c c c c c c c c c c}
\hline
{System} & {$a$ (\AA) } & {$b$ (\AA)} & {$\Delta_{bH}$ (\AA)} &  {width (\AA)} &  {$d_{C-C}$ (\AA)}  &  {$d_{C-H}$ (\AA)} & {$d_{C-F}$ (\AA)}  & $t$ (\AA) &  {C$_b$ (eVnm$^2$)} & N$_{\mathrm{chain}}$ & {F$_{zyzy}$ (nC/m)} \\[0.9ex]
\hline 
C & 2.47 & 4.278 & 0 &  77.006 & 1.43 &  & & 0.66 & 0.927  & 36 & 0.0004\\[1.5ex] 
CH & 2.54  & 4.40  &  0.46 &  79.12 & 1.54 & 1.11 &  & 2.68 & 1.68  & 36 & 0.035\\[1.5ex] 
CF & 2.60 & 4.50 & 0.49 &  81.06 & 1.58 & & 1.38  & 3.25 & 1.76    & 36  & 0.062\\[1.5ex] 
CHF & 2.58  & 4.47 & 0.47 & 80.44 & 1.56 &1.11 &1.39  & 2.96 & 1.43 & 36  & 0.167\\[1.5ex] 
\hline  
\end{tabular}

\end{table*}
Pure graphene, on the other hand, due to its centrosymmetric nature (point group D$_{6h}$) does not exhibit any piezoelectric properties. Several studies have focused on improving electro-mechanical properties of graphene by breaking its perfect planar symmetry in graphitic membranes~\cite{kalinin2008electronic}, carbon nanocones~\cite{kvashnin2015flexoelectricity}, carbon nanoshells~\cite{dumitricua2002curvature}, nanotubes~\cite{white1993predicting,artyukhov2020flexoelectricity}, graphene nanoribbon springs~\cite{javvaji2020exploration}, moir\'{e} superlattices~\cite{leovisualization} and by introducing various type of defects in perfect graphene sheet~\cite{javvaji2018generation}. However, in all the above mentioned studies the flexoelectricity is induced by symmetry breaking of $\pi$ orbitals under structural deformation which remains weak~\cite{kalinin2008electronic,kvashnin2015flexoelectricity,dumitricua2002curvature,javvaji2018generation}. Besides mechanical strains~\cite{ferralis2008evidence,kim2009large} and graphene origami~\cite{chen2019atomically, wei2020crease,ma2020multidimensional}, atomic functionalization is another popular experimental technique for engineering novel functionalities in graphene ~\cite{lee2012selective,johns2013atomic,elias2009control,pumera2013graphane, jorgensen2016symmetry, wheeler2012fluorine}. Previous reports have shown that the electronic properties of graphene can be modified by functionalization~\cite{paupitz2012graphene,jorgensen2016symmetry,yan2012chemistry}. Functionalization has also been used to induce piezoelectricity in graphene monolayer by adsorbing adatoms (such as H, F, Li) on a single side of the graphene monolayer~\cite{ong2012engineered}, or by co-adsorption of H and F~\cite{ong2013effect,kim2014origin}.

 Here, by performing state of the art first-principles calculations we propose that the rather inferior flexoelectric properties of graphene zig-zag nanoribbons (ZNRs) can be significantly improved in their hydrogen/fluorine functionalized variants $i.e$ in CH and CF nanoribbons. Due to the charge and ion redistribution upon bending, these ZNRs exhibits a net dipole moment which is linearly dependent on the applied strain gradient. We further show that the flexoelectric properties can be further improved in CHF type Janus ZNR by breaking both structural and charge symmetry. The bending energy of all ZNRs is small implying the feasibility of the proposed bending scheme. Furthermore, the electronic properties and hole effective masses can also be tuned by mechanical bending.

\begin{figure*}
\centering
\includegraphics[width=\textwidth]{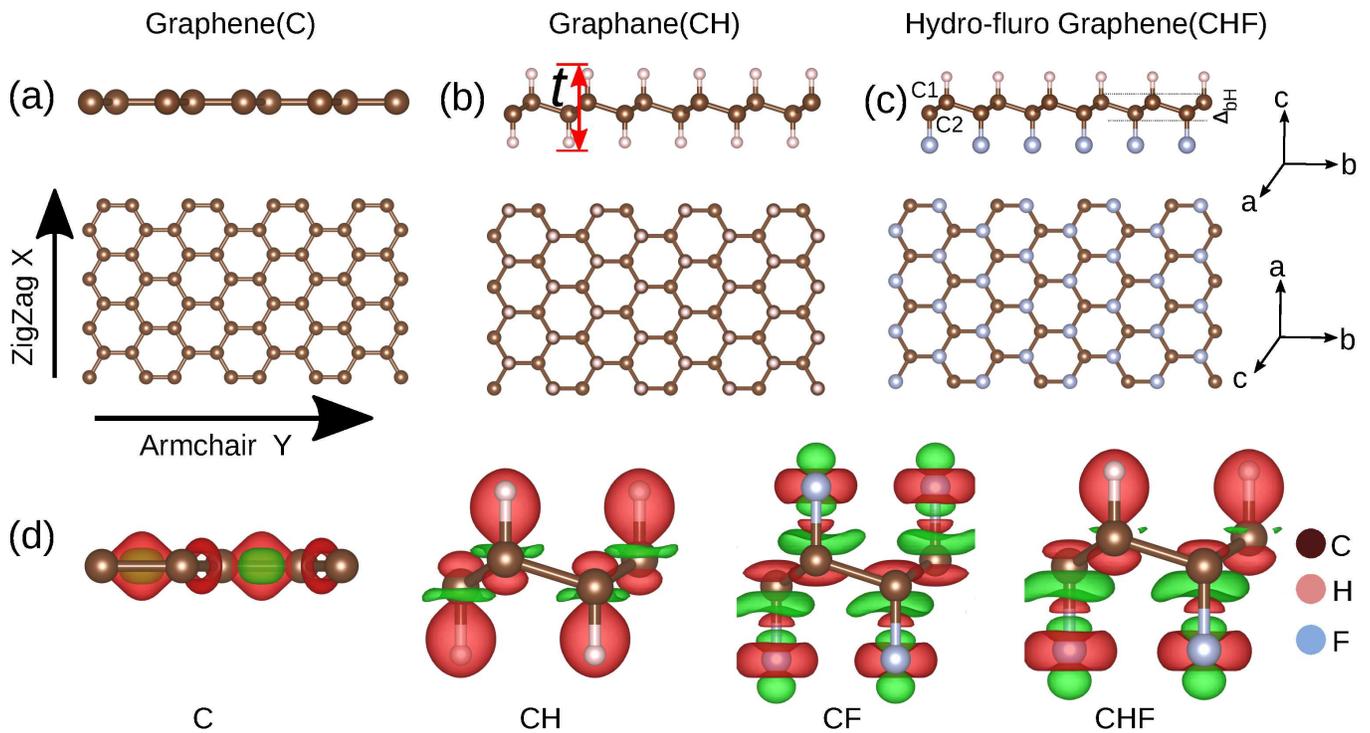}
\caption{Top and side view of monolayer structures of (a) C, (b) CH and (c) CHF. C1 and C2 represent carbon atoms (brown circles) connected to hydrogen (pink circles) and fluorine (cyan circles), respectively. The zig-zag and armchair directions are denoted as X and Y, respectively. The zig-zag nanoribbons were generated by including periodic boundary conditions along the $x$ direction and nanoribbon width is along the $y$ direction. $t$ in the part (b) is the distance between bottom and top H atoms described as the thickness of functionalized nanoribbons. (d) The charge density difference of C, CH, CF and CHF monolayers, plotted at an isosurface value of 0.013 e/~\AA$^3$. Red and green color present charge accumulation and depletion. The symmetric charge distribution of C-H and C-F bonds is broken in CHF monolayer, resulting in a induced dipole moment.}
\label{fig1}
\end{figure*}

\section{Theoretical and computational details}
We start from  zig-zag nanoribbons of graphene (C), graphane (CH), fluoro-graphene (CF) and hydro-fluoro-graphene (CHF). The lattice parameters for the corresponding monolayers and widths of the considered nanoribbons are listed in Table~\ref{Table:1}. The bent ZNR structures were created by displacing the $z$ positions of atoms with a quadratic function which depends on the $y$ position of atoms ($u^{z}= \frac{\kappa}{2} y^{2}$). Here $\kappa$ is the strain gradient of the bending plane defined as $\kappa = \frac{\partial^2 u^z}{\partial y^2}$. All the curved structures were generated by an in-house developed python workflow utilizing the Atomsk library~\cite{hirel2015atomsk}. For all ZNRs the bending is applied along $+ z$ direction. Additional details regarding the bending model employed here can be found in Refs.~\cite{javvaji2020exploration,zhuang2019intrinsic,hu2019atomistic}.

\begin{figure}
\centering
\includegraphics[width=\columnwidth]{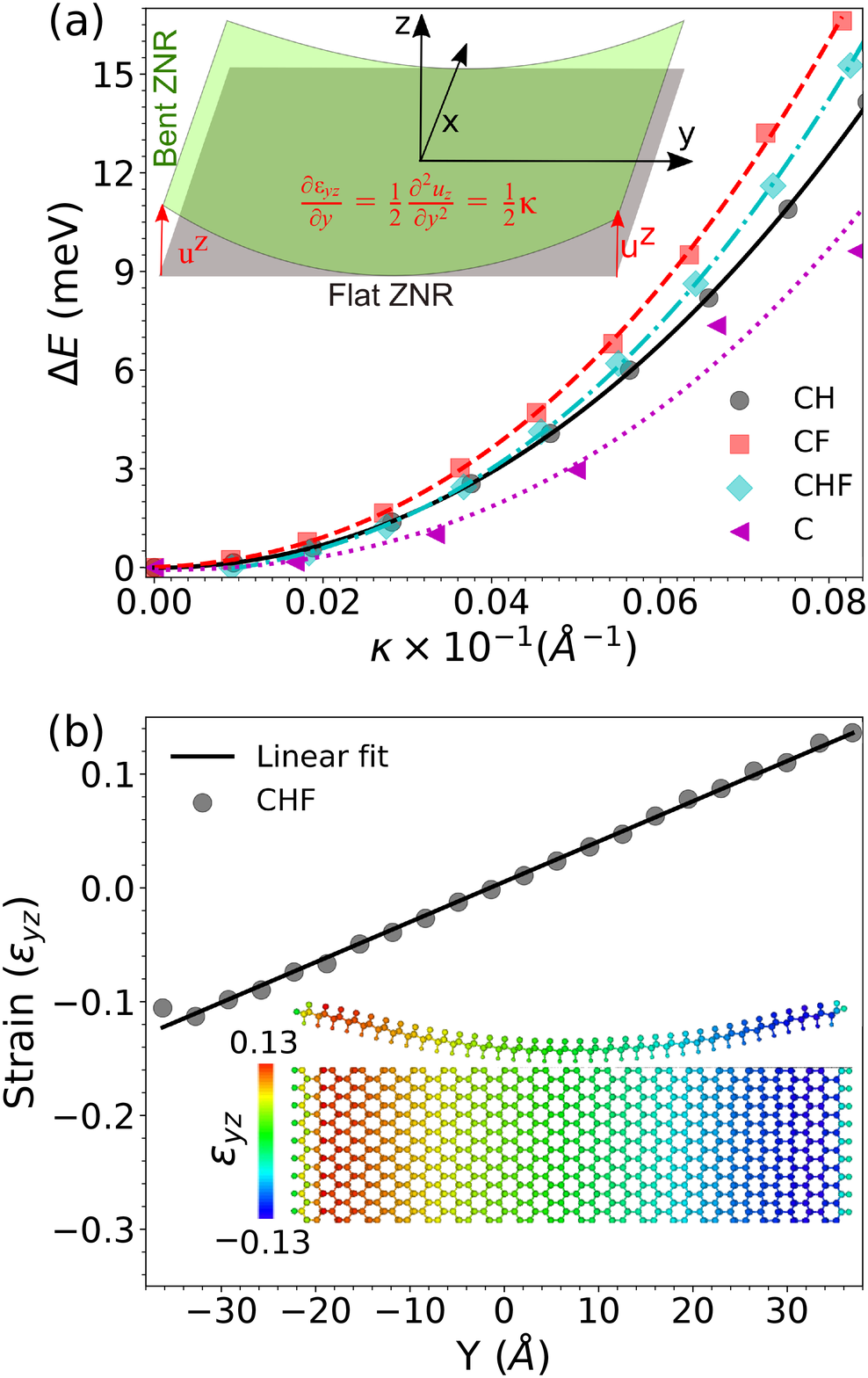}
\caption{(a) Strain energy as a function of applied strain gradient ($\kappa$) in C, CH, CF and CHF  zig-zag nanoribbon. The symbols represent the calculated values, while the solid lines are cubic polynomial fits. The inset in (a) shows a schematic of the applied bending scheme. (b) Variation of $\epsilon_{yz}$ with respect to Y atomic positions. The inset in (b) shows top and side view of induced atomic strain projected onto the atomic sites of CHF ZNR at a $\kappa$ value of 0.0073 \AA$^{-1}$.}
\label{fig2}
\end{figure}

These geometries were subsequently relaxed within density functional theory (DFT) simulations. All DFT calculations were performed within the Vienna \textit{Ab initio} Simulation Package (VASP)~\cite{Kresse} by using the projector augmented wave method with the Perdew-Burke-Ernzerhof (PBE) exchange correlation functional~\cite{PAW1,PAW2,PBE}. Periodic boundary conditions were applied in the $x$ direction to simulate the curvature of an infinitely long ribbon. A vacuum region of 25 ~\AA~ was used  along $y$ and $z$ directions. The edges of all ZNRs studied here were passivated with hydrogen. For the curved ribbons the edge atoms were kept fixed along $z$ and $y$ direction during structure relaxation while all interior atoms were fully relaxed by using a conjugate-gradient scheme until the force on the atoms were less than 0.01 eV/\AA. For flat ZNRs all atoms were relaxed. A 15 $\times$ 1 $\times$ 1 k-point grid was used to sample the Brillouin zone and energy convergence criteria of 1$\times$10$^{-5}$ was used during the structure relaxation. These calculations parameters and convergence criteria are standard and similar to those used in previous studies on mechanical properties of graphene and other related  2D materials~\cite{yu2016bending,nepal2019first}.

As discussed in the previous studies~\cite{zhuang2019intrinsic,hu2019atomistic} the applied bending perturbation only induces the $yz$ strain component and the strain gradient term ($\frac{\partial \epsilon_{yz}}{\partial \epsilon_{y}}$), whereas the remaining components of the strain and strain gradient tensors are zero~\cite{zhuang2019intrinsic,hu2019atomistic}. Therefore, the total induced polarization ($P_Z$) given by Eq. (\ref{eq1}) can be rewritten along the $z$-direction  as:
\begin{equation}
P_{Z} =  d_{zyz} \epsilon_{yz} + F_{zyzy}\frac{\partial \epsilon_{yz}}{\partial{y}},
\label{eq2}
\end{equation}

The dipole moment along the $z$ direction ($D_z$) was calculated with respect to the center of mass of the unit cell for all the systems. For dipole moment calculation a dense k-grid (51$\times$ 1 $\times$ 1) along with a high energy cutoff of 550 eV were used. Subsequently, the vertical polarization of the curved structure was estimated by $P_z$ = $D_z/V$, where $V$ is the effective volume of the flat ZNRs. Although the exact value of \emph{thickness} ($t$) for monoatomic layers such as graphene is debatable~\cite{wang2005size,huang2006thickness,kalinin2008electronic}, here we assume a thickness of 0.66~\AA~(the spatial extent of the graphene $p_z$ orbitals) for graphene ZNRs. For functionalized ZNRs thickness is defined as the distance between outer most atoms as described in the Figure~\ref{fig1} (b). The effective thickness employed in this work for all nanoribbons is listed in the Table ~\ref{Table:1}. This effective thickness was  used in calculations of effective volume of nanoribbons for describing the induced polarization.

\section{Results and discussions}
\subsection{Bending stiffness}
The relaxed structural parameters of C, CH, CF and CHF 2D layers corresponding to a rectangular unitcell are listed in Table~\ref{Table:1}. The most energetically favorable configuration for H and F co-adsorbed graphene monolayer (CHF structure in this work) was previously explored by Reed \textit{et. al.} ~\cite{ong2013effect,ong2012engineered}. They showed that the structure with hydrogen and fluorine being adsorbed on alternating atop carbon sites (the so called chair configuration) is energetically most favorable.  We use this configure for creating the CHF structure. The various structural parameters are highlighted in Figure~\ref{fig1}(a-c). Among these monolayers, graphene has the smallest and CF has the largest lattice parameter. Due to its higher electro-negativity, F induces larger buckling (0.49~\AA) than hydrogen (0.46~\AA). The induced charge distribution of all these monolayers is shown in Figure~\ref{fig1}(d). Even with complete hydrogenation or fluorination the centrosymmetry of the system is preserved and the charge distribution is symmetric. However, due to large electro-negativity difference H and F co-adsorption on graphene monolayer (CHF monolayer) breaks the symmetry, which results in a net dipole moment of 0.48 e\AA. Furthermore, the Janus CHF monolayer has asymmetric bond lengths and bond angles. As explained later this will have important consequences on out-of-plane flexoelectricty. These monolayers are repeated along the $x$ direction to create ZNRs with a finite width along the $y$-direction. The width of all the ZNRs studied here is listed in Table~\ref{Table:1}.

Next, we calculate the bending energy and bending stiffness of C, CH, CF and CHF ZNRs subjected to the bending perturbation described above. Within the strain gradient range explored here the bending energy of all ZNRs is smaller than 18.32 meV/\AA$^2$ --- the van der Waals binding energy of bulk graphite~\cite{lebegue2010cohesive,yu2016bending}. This small energy implies that the bending scheme employed in this work can be readily achieved via mechanical bending or can be introduced during growth. The bending stiffness ($C_b$) is calculated by taking the second derivative of the strain energy on a per atom basis (energy difference between the flat and the bent ZNRs divided by the number of atoms) with respect to the strain gradient ($\kappa$) as shown in Figure~\ref{fig2}(a). The lowest bending stiffness of 0.927 eVnm$^2$ is obtained for C ZNRs, followed by CH, CHF and CF ZNRs. Note that the bending stiffness values obtained here are higher than the previously reported values~\cite{mintmire1995electronic, kudin2001c, gonzalez2018bending} for graphene, where the bending scheme was related to the generation of carbon nanotubes. However, the bending scheme employed in this work displaces the atoms only along the $z$ direction resulting in higher perturbation of bond lengths and therefore a larger bending stiffness is obtained.
\begin{figure}
\centering
\includegraphics[width=\columnwidth]{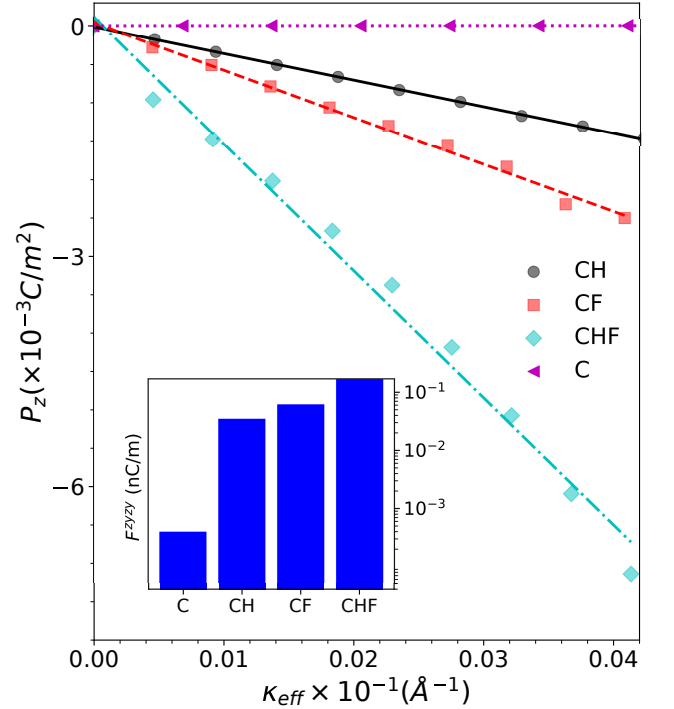}
\caption{Induced out of plane polarization ($P_z$) vs curvature for C, CH, CF and CHF zig-zag nanoribbons. Symbols are calculated values and lines are linear fit. The inset gives the out of plane flexoelectric coefficient ($F^{zyzy}$) derived from  the slope of the linear fit. Note the log y-scale in the inset. For the case of CHF, the flat nanoribbon also has finite non-zero dipole moment (polarization) and the induced polarization plotted is with respect to flat nanoribbon.}
\label{fig3}
\end{figure}

\subsection{Bending induced strain and flexoelectricity}

As alluded before the bending model employed here induces a strain in $yz$ direction and a strain gradient $\frac{\partial \epsilon_{yz}}{\partial y}$ in the $yz$ direction and that in-turn gives rise to induced polarization along the $z$ direction. Figure~\ref{fig2}(b) shows the linear dependence of 
$\epsilon_{yz}$ with respect to $y$ atomic positions for CHF ZNRs at 0.0073 \AA$^{-1}$ strain gradient. The inset of Figure~\ref{fig2}(b) presents the $\epsilon_{yz}$ strain component projected onto CHF atomic structure. The atomic strain at site $i$ is computed according to~\cite{li2005least,shimizu2007theory, stukowski2009visualization}
\begin{equation}
\epsilon_i = \frac{1}{2}[(F_i)^{T} F_i -I],
\end{equation}
where $F$ is the deformation strain gradient tensor and I is the identity matrix. As reported previously~\cite{zhuang2019intrinsic, javvaji2019high} the strain induced in the $yz$ directions varies linearly, indicating the symmetric nature of induced deformation. Therefore, any contribution to the induced polarization due to $\epsilon_{yz}$ in Eq.~\ref{eq2} is eliminated. 

Accordingly, the flexoelectric response ($F_{zyzy}$) is described as the induced polarization due to a strain gradient, which in this case can be described as~\cite{zhuang2019intrinsic,hu2019atomistic,javvaji2019high}:
\begin{equation}
P_z= F_{zyzy}\frac{\partial \epsilon_{yz}}{\partial y}
\end{equation}
From the slope of Figure~\ref{fig2}(b) we can obtain $\kappa_{\mathrm{eff}}$ ($\frac{\partial \epsilon_{yz}}{\partial y}$), which includes the effect of structural relaxation on the applied strain gradient. For instance, the slope of  Figure~\ref{fig2}(b) gives $\kappa_{\mathrm{eff}}$ as 0.0035~\AA$^{-1}$, which is  $\sim$ 3.5\% smaller than the actual applied $\kappa/2$. Similarly, we find that for all the systems studied here $\kappa_{\mathrm{eff}}$ is 2-5\% smaller than the initial applied $\kappa/2$. This $\kappa_{\mathrm{eff}}$ is subsequently used in the calculation of the flexoelectric coefficients.

\subsubsection{Flexoelectricity in C, CH and CF zig-zag nanoribbons}
The estimated induced polarization (dipole moment per unit volume) is plotted in Figure~\ref{fig3} as a function of $\kappa_{\mathrm{eff}}$. As expected, with increasing $\kappa_{\mathrm{eff}}$, due to higher non-linear strain gradients, the magnitude of induced polarization also increases. The out of plane flexoelectric coefficient can then be obtained by linearly fitting the induced polarization, which is plotted in the inset of Figure~\ref{fig3}. The lowest flexoelectric coefficient was observed for graphene ZNRs, followed by CH and CF ZNRs, respectively. For graphene ZNRs the calculated flexoelectric coefficient is 0.0004 nc/m. Initially, for the flat graphene ZNR the out of plane dipole moment is zero, however as the graphene ZNR is bent the $\pi$-$\sigma$ interactions begin to increase, which induces a net non zero dipole moment along $z$ direction. To analyse the effect of strain gradient on the induced polarization, the electronic and ionic components of the dipole moment were calculated within the Berry phase method~\cite{spaldin2012beginner,vogl1978dynamical,resta1994macroscopic}. We find that for graphene nanaribbons with increasing $\kappa_{\mathrm{eff}}$ the ionic component of the dipole moment exhibits a larger change than the electronic part of the dipole moments. Moreover, as the $\kappa_{\mathrm{eff}}$ increases, the contribution from the electronic component becomes much more important. For example, for a strain gradient of 0.002 \AA$^{-1}$, while  93.4 \% of the total dipole moment has an ionic origin, only 5.6 \% has an electronic one. When the $\kappa_{\mathrm{eff}}$ increases to 0.004 \AA$^{-1}$ the contribution of ionic and electronic component of the dipole moments becomes $\sim$ 91.1 \%  and 7.9 \%, respectively. The increase in electronic dipole moment with bending can be attributed to the enhanced $\pi$-$\sigma$ interactions originating from strain gradient driven $sp^2$ to $sp^3$ hybridization transition ~\cite{dumitricua2002curvature,kundalwal2017strain}. This is also in agreement with a recent study on graphene where the polarization was calculated from a charge-dipole model~\cite{zhuang2019intrinsic,kundalwal2017strain}. 

\begin{figure}
\centering
\includegraphics[width=\columnwidth]{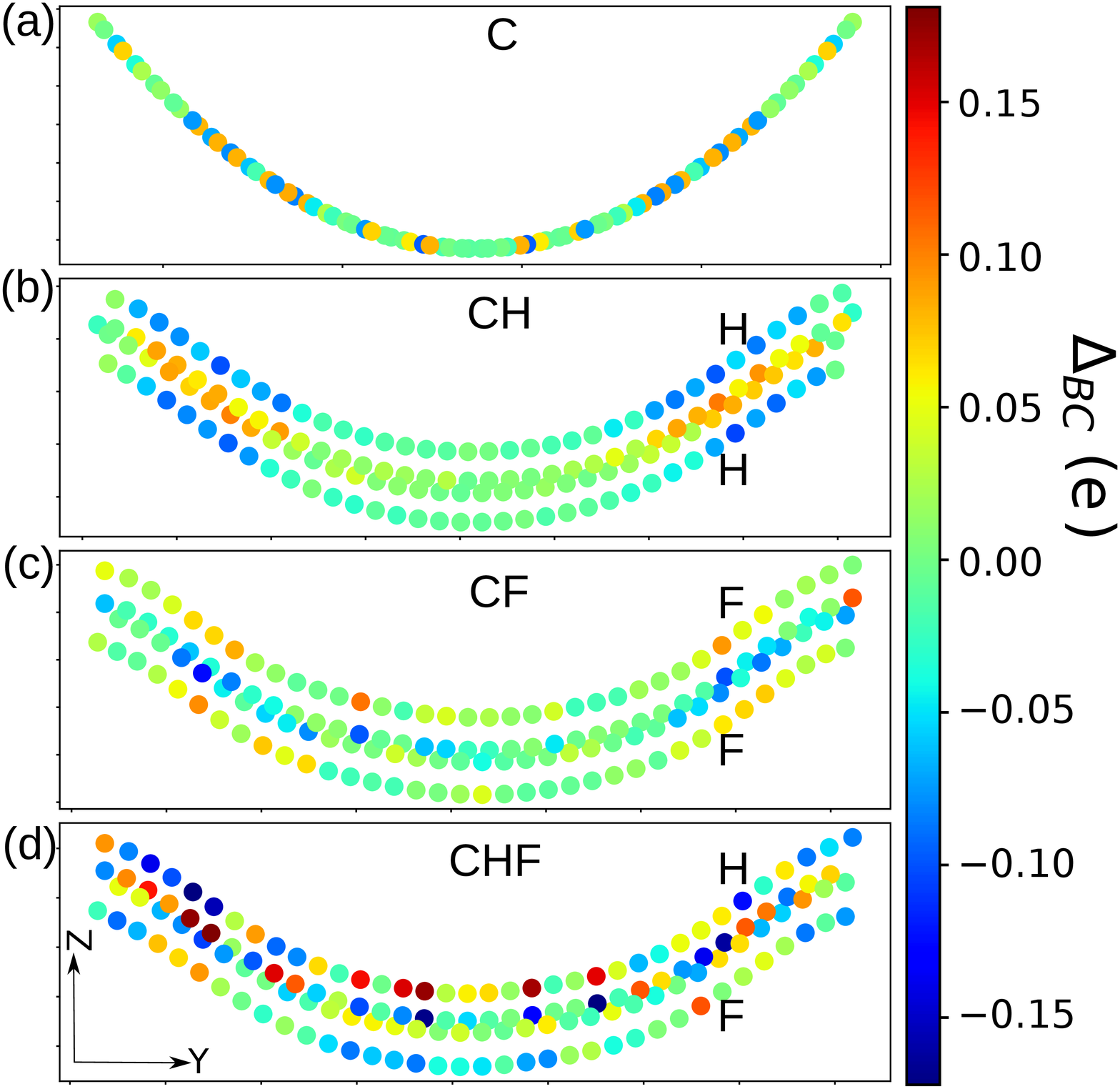}
\caption{Redistribution of Bader charges ($\Delta_{BC}$ in the unit of electron charge) with respect to flat ribbons at a $\kappa_{\mathrm{eff}}$ of 0.0041 \AA$^{-1}$ for (a) C, (b) CH, (c) CF and (d) CHF. As can be seen the symmetries of charges is broken under the applied strain gradient.}
\label{fig4}
\end{figure}

\begin{figure}
\centering
\includegraphics[width=\columnwidth]{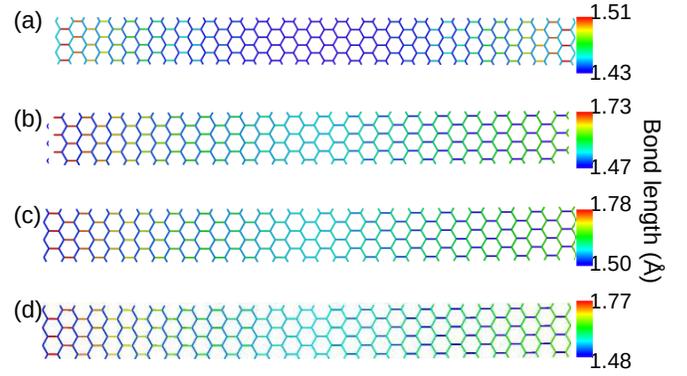}
\caption{Effect of applied displacement field ($\kappa_{\mathrm{eff}}$ = 0.0041 \AA$^{-1}$) on the C-C bonds in (a) C, (b) CH, (c) CF and (d) CHF. For comparison the C-C bond length in the flat C, CH, CF and CHF ribbons are 1.426, 1.51, 1.52 and 1.53 \AA, respectively. Upon bending the equal C-C bonds in the graphene unit cell become different. At the same value of $\kappa_{\mathrm{eff}}$ a maximum and minimum variation in bond lengths is observed in CHF and C, respectively.}
\label{fig5}
\end{figure}
 
As shown in Figure~\ref{fig3} the induced polarization for CH and CF ZNRs is much larger than that in graphene ZNRs.  The higher induced polarization results in larger $F^{zyzy}$ values of 0.035 nC/m and 0.062 nC/m for CH and CF ZNRs, respectively. Structurally there are two main differences between C and CH/CF ZNRs; (i) H or F atoms induce buckling (0.46-0.49 \AA) in the flat graphene sheet and (ii) the out of plane CH bonds provide additional avenues for tuning the charge transfer under mechanical bending. As previously noted for buckled silicene~\cite{zhuang2019intrinsic} such structural buckling is useful for generating  higher flexoelectricity.

To investigate the effect of H/F atoms on flexoelectric properties we next analyse the Bader charges~\cite{sanville2007improved,tang2009grid}. Initially, for the flat ($\kappa_{\mathrm{eff}}$ = 0) CH and CF ZNRs, the Bader charges of both top and bottom H ($-$0.11e) and F ($-$0.52e) are  equal. As the bond lengths between top and bottom H (F) and C atoms are the same, the dipole moment of C-H$_{\mathrm{bottom}}$ (C-F$_{\mathrm{bottom}}$) bond cancels the dipole moment of other C-H$_{\mathrm{top}}$ (C-F$_{\mathrm{top}}$) bond, resulting in a net zero total dipole moment. The charge symmetry of these top and bottom C-H or C-F bonds is perturbed under applied strain gradient. The changes in Bader charges with respect to flat ZNRs at $\kappa_{\mathrm{eff}}$ of 0.0041\AA$^{-1}$ are shown in Figure~\ref{fig4}. 

As can be seen, the bending modifies the bond lengths (shown in Figure~\ref{fig5}), therefore resulting in charge redistribution within the ZNRs. The magnitude of this charge redistribution is directly proportional to $\kappa_{\mathrm{eff}}$. For example, in graphene ZNRs (Figure~\ref{fig4}) at $\kappa_{\mathrm{eff}}$ = 0.0041 \AA$^{-1}$, the charge redistribution varies between $-$0.075\textit{e} and 0.075\textit{e}, resulting in a finite induced dipole moment. When bending is applied to the CH or CF ZNRs, along with C-C bond lengths, the out of plane C-H and C-F bond lengths also change. The C-C bond lengths at $\kappa_{\mathrm{eff}}$ = 0.0041 \AA$^{-1}$ are shown in Figure~\ref{fig5}(a-d). The larger bond length perturbation results in a larger redistribution of charges under bending. For instance, at the same $\kappa_{\mathrm{eff}}$ (0.0041 \AA$^{-1}$) for CF monolayer the change in the Bader charges varies from $-$0.15e to 0.15e, which is much larger than in the graphene ZNR. Therefore, for similar $\kappa_{\mathrm{eff}}$, higher induced polarization is observed in CH and CF ZNRs, implying larger out of plane flexoelectric coefficients.
\begin{figure}
\centering
\includegraphics[width=\columnwidth]{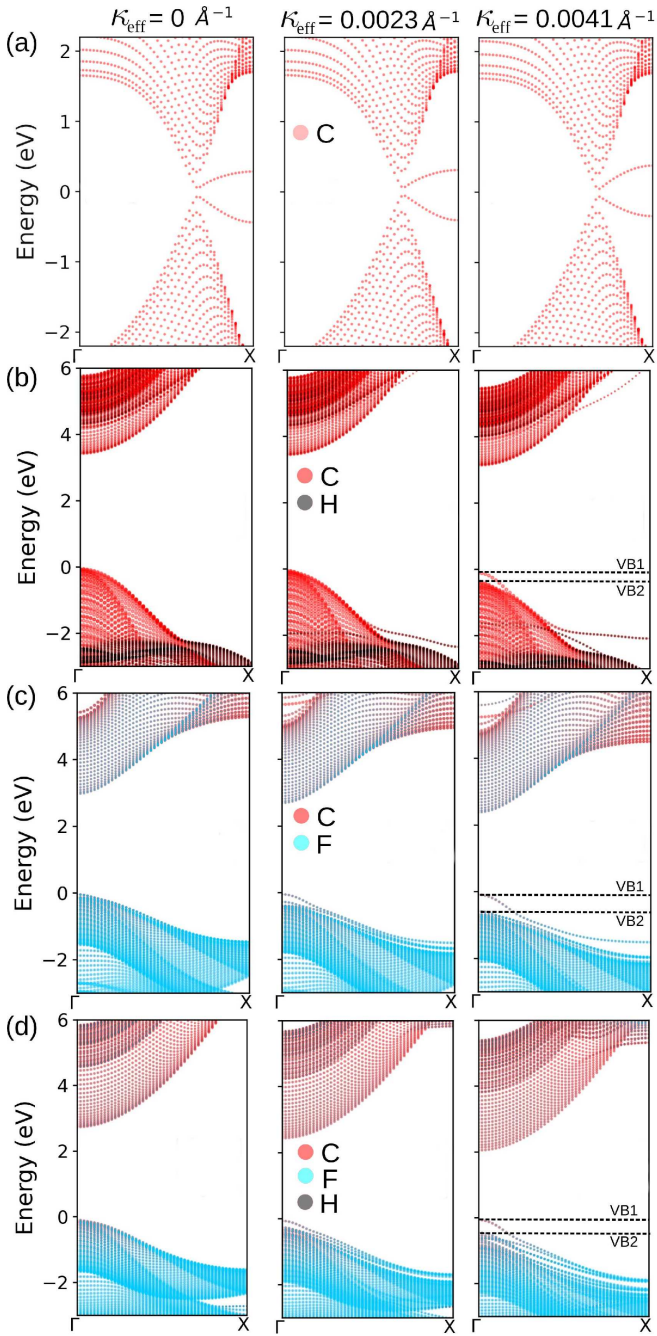}
\caption{Bending effects on the electronic structure of zig-zag (a) C, (b) CH, (c) CF and (d) CHF ZNRs at selected effective strain gradient ($\kappa_{\mathrm{eff}}$). The atomic projected eigenvalues are shown in different colors.}
\label{fig6}
\end{figure}

\subsubsection{Flexoelectricity in CHF zig-zag nanoribbons}
We next explore the possibility to enhance the flexoelectric properties of graphene ZNRs by dual functionalization \textit{i.e.} creating CHF type ZNRs as shown in Figure~\ref{fig1}(c). Due to the large electro-negativity difference between H (2.2) and F (3.98); the CH and CF bond lengths in a CHF monolayer are very different (Table~\ref{Table:1}). The charge density difference in a CHF monolayer is shown in Figure~\ref{fig1}(d). As can be seen from  Figure~\ref{fig1}(d) for CHF monolayer the charge symmetry of C-H and C-F bonds is broken. The Bader charge analysis shows that both H ($-$0.11e) and F ($-$0.54e) atoms in the flat CHF ZNRs are negatively charged. Whereas, the averaged Bader charge on the C atoms connected to H is 0.11e and for those connected to F is 0.54e. Therefore, the local dipole induced in C-F bonds dominants over the C-H bonds, resulting in a net dipole moment. When the CHF ZNR is bent, the charge asymmetry in the buckled carbon layer is further increased. Due to the large electro-negativity difference between F and H atoms, the C1 atom (the carbon atom attached to H, as marked in \ref{fig1}(c)) and C2 atoms (carbon atom bonded to F) experience different perturbation, resulting in different charges on these atoms. When the ZNR is bend  ($\kappa_{\mathrm{eff}}$ = 0.0041 \AA$^{-1}$) the charge in the CHF ZNR at the C1 site is enhanced by  20\%, while the charge at the C2 site decrease by 12\%.  The corresponding number for CF ZNRs at C1 and C2 sites are 5\% and 3\%, respectively at the same $\kappa_{\mathrm{eff}}$. Therefore, for the same $\kappa_{\mathrm{eff}}$ CHF ZNRs have higher induced polarization, implying an even larger flexoelectric coefficient. From the linear fitting of induced polarization our calculations predict a F$_{zyzy}$ of 0.167 nC/m, which is nearly 2.5 times larger than the  F$_{zyzy}$  value for CF.

For CH and CF ZNRs due to the identical C-H or C-F bond lengths at the flat state the bending along $\pm z$ direction is equivalent. However in the case of CHF ZNRs, C-F bond lengths are larger than C-H bond lengths, therefore bending along $+$ or $- z$ direction may be energetically different. To investigate this further we compare the bending energy of CHF ZNRs along $+z$ (bending is applied towards H atoms) and $-z$ direction (bending is applied towards F atoms). We find that, bending along the $+z$  direction (along C-H bonds) is energetically favorable and the energy difference relative to $-z$ direction is $\sim$ 510 meV per unit cell (3.53  meV per atom basis) at $\kappa_{\mathrm{eff}}$ = 0.0041 ~\AA$^{-1}$. Nonetheless, we computed the induced polarization of CHF ZNRs under bending along $-z$ direction and find a F$_{zyzy}$ of 0.154 nC/m. This value is comparable to the 0.167 nC/m, which is obtained for bending along $+ z$ direction. 

Furthermore, to investigate the effect of the width of ZNRs on flexoelectric properties, additional calculations were performed for larger width CH and CHF ZNRs  (w = 182.374 \AA~for CH and w = 184.87 \AA~for CHF). For these larger width CH and CHF ZNRs the calculated  F$_{zyzy}$ are 0.038 nC/m and 0.175  nC/m, respectively. These F$_{zyzy}$ values are comparable (within 8 \%, and 5 \% for CH and CHF respectively) to the values computed with the smaller widths ZNRs listed in Table~\ref{Table:1}. Therefore, the width of ZNRs simulated here is likely large enough to eliminate finite size effects on the flexoelectric properties.

\begin{figure}
\centering
\includegraphics[width=\columnwidth]{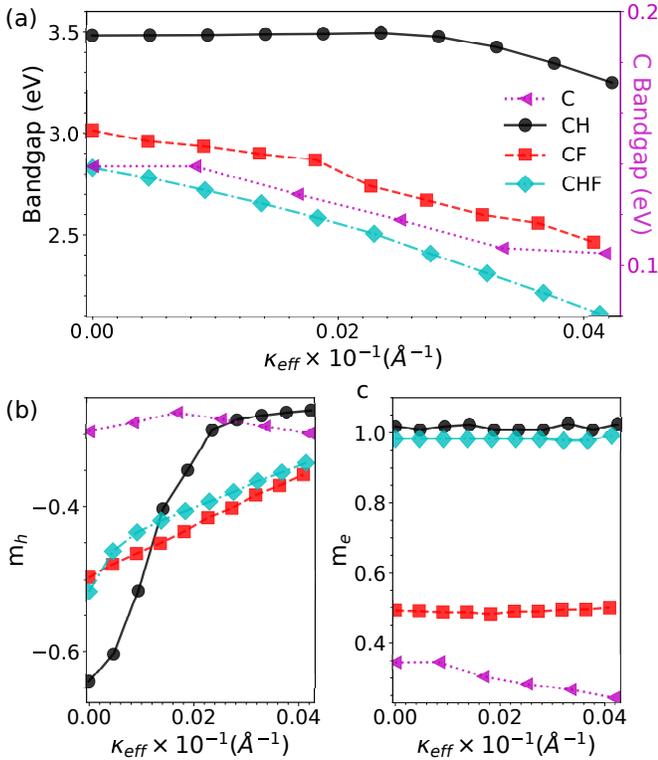}
\caption{(a) Calculated band gaps as a function of effective strain gradient ($\kappa_{\mathrm{eff}}$) for C, CH,CF and CHF  zig-zag nanoribbon. The band gap of C nanoribbons is shown on right y-axis. (b) Variation in the hole ($m_h$) and (c) electron effective mass ($m_e$) (in the unit of bare electron mass) along the $x$ direction as function of $\kappa_{\mathrm{eff}}$.}
\label{fig7}
\end{figure}

\subsection{Effect of bending on the electronic properties}
Mechanical bending also impacts the electronic properties as it can induce localization/delocalization in the electronic charges and modify hybridization. This change in the charge distribution leads to modifications of electronic properties such as the effective mass and band gap. The effect of bending on C, CH, CF and CHF ZNRs band structures is shown in Figure \ref{fig6}(a-d), respectively at selected $\kappa_{\mathrm{eff}}$ values. Here atomic projected eigenvalue values are also shown in different colors. The variation of band gap as a function of  $\kappa_{\mathrm{eff}}$ is shown in Figure \ref{fig7}(a). Bending has no significant effect on the electronic properties  of graphene ZNRs and with bending the band gap reduces by 0.032 eV as shown in Figure \ref{fig7}(a). The electronic structure of graphene is mainly sensitive to uni-axial in plane strain, twisting~\cite{tang2012altering} The bending scheme employed here induces a small in-plane strain, which is insufficient to induce any noticeable hybridization changes. This can also be co-related with the strain gradient induced polarization which is small in the case of C ZNRs. The effect of a strain gradient on C ZNRs hole ($m_h$) and electron $m_e$ effective masses is shown in Figure~\ref{fig7}(b) and (c) and both $m_h$ and $m_e$ mostly remain unchanged. 

The situation is different for CH, CF and CHF ZNRs, where electronic properties are sensitive to C-H, or C-F out of plane bonds as well. For the flat CH ZNRs the band gap is about 3.47 eV, where both VBM and CBM originate mainly from carbon atoms as shown in Figure~\ref{fig6}(b). As the CH ZNR is bent, initially the band gap starts to decrease at a slower rate and as the bending exceeds 0.0025 \AA$^{-1}$ the band gap decreases sharply as shown in Figure~\ref{fig7}(a). Overall, within the $\kappa_{\mathrm{eff}}$ range explore here the band gap of CH ZNRs is reduced by $\sim$ 0.3 eV. In the case of CF and CHF ZNRs the band gap drops at a much faster rate as a function of bending. This is because for CH ZNRs the valence band maxima, originates from C atoms, however for CF and CHF ZNRs the VBM shows a strong hybridization between C and F atoms. When mechanical bending is applied, the distance between C-F atoms increases, which reduces the hybridization between C-F states, leading to large splitting between the VB1 and VB2 states as shown in Figure~\ref{fig6}(b) and (c). The bending not only reduces the band gap of CH, CF and CHF ZNRs but also reduces the hole effective masses as shown in Figure~\ref{fig7}(b), whereas the electron effective mass remains mostly unchanged (Figure~\ref{fig7}(c)). 

\begin{figure}
\centering
\includegraphics[width=\columnwidth]{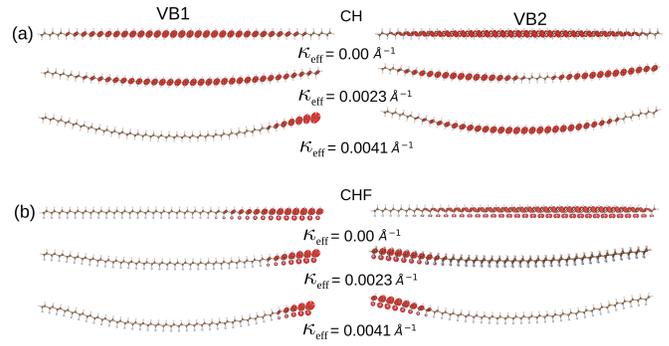}
\caption{Variation in the iso-surface of partial charge densities for valence bands at the VB1 and VB2 under $\kappa_{\mathrm{eff}}$ for (a) CH and (b) CHF. The partial charge density of CF nanoribbons follows the same trend as CHF nanoribbons with $\kappa_{\mathrm{eff}}$. These are plotted at the iso-surface value of 0.0005 eV/\AA$^3$. VB1 and VB2 are marked in Figure~\ref{fig6}.}
\label{fig8}
\end{figure}

The variation of valence band partial charge density at VB1 and VB2 at $\Gamma$-point (marked in Figure~\ref{fig7}) is shown in Figure~\ref{fig8}(a)-(b) for CH and CHF, respectively. Using mechanical bending as an external perturbation the distribution of these charges can be tuned together with the electronic properties~\cite{yu2016bending,nepal2019first}. In the case of CH ZNRs at zero $\kappa_{\mathrm{eff}}$ the charge carriers in both VB1 and VB2 states are delocalized over the entire ribbon width. With the increase in $\kappa_{\mathrm{eff}}$ the charges of VB1 bands are localized towards the right edge of the ZNR. At the same time the charge density more or less remains distributed over the entire ribbon width for VB2. A different trend is observed for CF and CHF ZNRs. As shown in Figure~\ref{fig8}(b), even at zero $\kappa_{\mathrm{eff}}$ the partial charges in the CHF ZNRs are delocalized along the right edge and with increasing $\kappa_{\mathrm{eff}}$  this edge charge delocalization becomes more prominent. Also, the charge in the VB2 state are more delocalized along the left edge. The large variation in the hole effective masses under $\kappa_{\mathrm{eff}}$ for CH ZNRs can be explained on the basis of this partial charge density. When charges are distributed over the entire ribbons width the hole effective masses are larger. With increasing $\kappa_{\mathrm{eff}}$ the charges get accumulated towards right edge the effective mass decreases. At higher $\kappa_{\mathrm{eff}}$, for VB1, all three ZNRs exhibit similar partial charge density and therefore the hole effective masses are comparable.

The dependence of bandgap on the ZNR width and number of zigzag chains is also studied. This is shown in the supplementary material Figure S1, along with the band structures of corresponding monolayers (Figure S2). As can be seen with the increasing width, the bandgap of all ZNRs decreases and converges to the bandgap value of corresponding monolayers. For graphene ZNRs our results are in good agreement with previous calculations of Son  \textit{et. al.} ~\cite{PhysRevLett.97.216803}.

\section{Discussion and Summary}
Our extensive first-principles calculations unveil a significant out of plane flexoelectricity in functionalized graphene ZNRs which is comparable to the recently reported values for monolayer transition metal dichalcogenides~\cite{javvaji2019high,zhuang2019intrinsic}. We predict that flexoelectricity in hydro-fluorinated graphene  ZNRs (CHF) is three times larger than that of CF ZNRs. This large enhancement in flexoelectricity is mainly facilitate by the large electro-negativity difference between H and F which breaks the structural and charge symmetry. By analyzing the electronic structure as a function of $\kappa_{\mathrm{eff}}$ we find that for the functionalized ZNRs the hole effective mass can be reduced by $\sim$ 10\%, whereas the electron effective mass mostly remains unchanged.

Previous reports have shown that single side H or F functionalization can induced strong piezoelectricty in graphene~\cite{paupitz2012graphene,jorgensen2016symmetry,yan2012chemistry} and silicene~\cite{noor2015hydrogen} monolayers. These could naturally also be an interesting avenue for tuning the flexoelectric properties. Our calculations for single side H functionalized graphene ZNR, predict a F$_{zyzy}$ value of 0.053 nC/m which is higher than the value (0.035 nC/m) obtained for double side functionalized graphene (CH) ZNR. Although, we do not explicitly calculate properties for single side F functionalized C ZNRs, based on the trend observed for H functionalized ZNRs, relatively higher flexoelectric properties can be anticipated. Similarly, oxygen functionalized graphene ZNRs are also expected to show good flexoelectric properties. Moreover the ever growing interest in Janus 2D materials where intrinsically a local dipole moment is present due to symmetry breaking could provide an exciting platform for low dimensional flexoelectricity.

The main highlight of the above results is that the flexoelectric properties of many low dimensional materials could be significantly improved by simple functionalization, which is readily achievable in experiments. Additionally, the bending scheme proposed here could be realized via mechanical bending, or can be introduced during the growth. These results open new opportunities to take advantage of the unique mechanical bending of low dimensional materials not only to investigate their flexoelectric properties, but also to tune their physical properties such as band gap and effective masses. With the recent experimental advances in the growth of graphene nanoribbons ~\cite{kolmer2020rational}, we believe that our findings will stimulate future experimental exploration of flexoelectric properties of these and other similar compounds.

\section*{Acknowledgment}
The computational resources and services used for the first-principles calculations in this work were provided by the VSC (Flemish Supercomputer Center), funded by the Flemish Science Foundation (FWO-VI). T. P. is supported by a postdoctoral research fellowship from BOF-UAntwerpen. 

%

\end{document}